\documentclass[aps,prd,preprint,tightenlines,
   noshowpacs,nofootinbib]{revtex4}
\newcommand{\PRE}[1]{{#1}}   

\usepackage{bm}
\usepackage{epsfig}
\let\d=\delta
\let\e=\epsilon\let\g=\gamma

\newcommand{\eqref}[1]{Eq.~(\ref{#1})}

\newcommand{\be}{\begin{equation}}
\newcommand{\ee}{\end{equation}}
\newcommand{\bea}{\begin{eqnarray}}
\newcommand{\eea}{\end{eqnarray}}

\newcommand{\del}{\partial}

\begin{document}

\preprint{hep-th/0405024} \vskip 1 cm
 \preprint{ MCTP-04-26}

 \preprint{
UCI-TR-2004-15}

\title{
\PRE{\vspace*{1.5in}} Revisiting
D-branes in $AdS_3 \times S^3$ \PRE{\vspace*{0.3in}} }

\author{ Jason Kumar
\footnote{email address: jpkumar@umich.edu }}
\affiliation{Michigan Center for Theoretical Physics, University of
Michigan Ann Arbor, MI  48105 USA \PRE{\vspace*{.3in}} }
\author{Arvind Rajaraman
\footnote{email address: arajaram@uci.edu}}
\affiliation{Department of Physics and Astronomy, University of
California, Irvine, CA 92697, USA \PRE{\vspace*{.3in}} }


\begin{abstract}
\PRE{\vspace*{.1in}} We find the most general supergravity solution
in an $AdS^3 \times S^3$ background preserving an $AdS_2 \times S^2$
symmetry and half the supersymmetries. Contrary to previous
expectations from boundary state arguments, it is shown that no
solutions exist containing localized brane sources.
\end{abstract}

\maketitle

\section{Introduction}

The AdS/CFT correspondence has led to a much deeper understanding of
nonperturbative string theory. However, most analyses of the
correspondence have been restricted to the supergravity
approximation. It is important to understand more about the truly
stringy aspects of the theory.

Major progress was made recently in understanding the quantization of
strings in these backgrounds \cite{mo}. Much of this progress relies
on the fact that the $AdS_3 \times S^3 \times T^4$ background with
NS-NS fluxes can be described as a Wess-Zumino-Witten (WZW) theory,
where the $AdS_3 $ factor corresponds to the noncompact group
manifold $SL(2,R)$, and the sphere to a compact group manifold
$SU(2)$. This allows the classification of the string states in terms
of representations of these groups.

 One can also attempt to construct D-branes in $AdS_3 \times
S^3$ using the method of conformal field theory boundary states. This
was done for the case of compact groups long ago, and the D-branes
can be classified. However, the noncompact case is much more subtle.

The basic idea is to treat the D-brane as a boundary state
$|B\rangle$ satisfying the condition $(J+J^\dagger) |B\rangle=0$.
(Here $J$ is the worldsheet current of the theory.) This can be
solved in the compact case in terms of Ishibashi states
\cite{Ishibashi:1988kg}. The D-branes are then linear combinations of
Ishibashi states that satisfy the Cardy condition \cite{Cardy:ir}.
For  compact groups, such states can be constructed in a
well-understood way. The construction of similar boundary states for
SL(2,R) was attempted in the papers \cite{Rajaraman:2001cr};
unfortunately, there may be problems with this construction
\cite{Ponsot:2002te}.

We shall study this issue here from another angle, by trying to
construct supergravity solutions for D-branes. The boundary state
condition implies by semiclassical reasoning \cite{Bachas:2000fr}
that the D-branes must be one-half BPS i.e. they preserve 8 of the 16
supersymmetries of the background. Furthermore, they must wrap $AdS_2
\times S^2$ submanifolds of the total space. We look for such branes
by explicitly analyzing the Killing equations, and finding the most
general solutions.

Our results are surprising: even at the linearized level, we show
that there are no localized branes with $AdS_2 \times S^2$ geometry
that preserve half the supersymmetries. There are hence no such
half-BPS branes in the $AdS_3 \times S^3$ background. We do find a
large class of delocalized solutions, some of which were also
previously discussed in \cite{Kumar:2002wc,Kumar:2003xi}. We can
write down the solutions to the nonlinear Killing equations in terms
of a few functions.

After first reviewing our notation in section II, we present the
linearized analysis of the Killing equations in section III and show
that consistency requires the charge to be delocalized. We then move
to the full nonlinear analysis in section IV and present the most
general solution (with delocalized charges) that preserves half the
supersymmetries of the $AdS_3\times S^3$ space. We close with some
comments on the relation between this calculation and the boundary
state construction in section V.

\section{Calculating the Solution}

We will consider solutions to Type $IIB$ supergravity, using the
conventions of \cite{deHaro:2002vk}.  The relevant bosonic field
content will consist of the vielbein, complex 3-form field strength $
f_{3}$, real 5-form field strength $f_{5}$, the complex 1-form
dilaton-axion
field strength $P$, and the composite connection $Q$. We will
consider perturbations around the $AdS_3 \times S^3 \times T^4$
background with NS-NS fluxes turned on.  For this background, the
metric can be written in the form  \bea ds^2 =(\bar{e}_{m \mu }
dx^{\mu })(\bar{e}^{m}_ {\nu }dx^{\nu})~~~~~~~~~~~~~~~~~~~~~
~~~~~~~~~~~~~~~~~~~~~~~~~~~~~~~~~~~~~~~~~~~~~~~~
~~~~~~~~~~~~
\nonumber\\
= -d\psi ^2 -\cosh^2 \psi (d\omega^2 -\cosh^2 \omega d\tau ^2)
-d\theta^2 -\sin^2 \theta (d\phi^2 +\sin^2 \phi d\chi^2 )
-(dx_a)(dx^a)\eea Here the indices $\psi, \tau, \omega$ parametrize
the $AdS_3$, $\theta,\phi,\chi$ parametrize the $S^3$ and $x_a
(a=1..4)$ parametrize the $T^4$. The background $NS$-$NS$ field
strengths are \bea  f^{\psi \tau \omega} &=&  f^{\phi \chi \theta} =1
\eea (we shall use indices with a tilde to denote curved-space
indices, and indices without a tilde to refer to tangent space
coordinates.)

This solution preserves 16 supersymmetries, which are encoded in the
Killing spinor $\tilde{\e}$. These Killing spinors are subject to the
constraint $\gamma_{\psi \tau \omega \phi \chi \theta} \tilde
\epsilon = \tilde \epsilon$ and satisfy the Killing equation \bea
\del_\mu \epsilon -{1\over 4} \omega_{ \mu} ^{~pq} \gamma_{pq}
\epsilon -{1\over 4} Q_\mu \e &-&{\imath \over 192}  f_{\mu pqrs}
\gamma^{pqrs} \epsilon -{3 \over 16}  f_{\mu pq} \gamma^{pq}
\epsilon^*
+ {1 \over 48}  f^{pqr} \gamma_{\mu pqr} \epsilon^*= 0
\eea

In addition, the vanishing of the variation of the dilatino gives the
equation \bea {1 \over 2}p_\mu  \gamma^{ \mu} \epsilon^* +{1 \over
24} f^{abc} \gamma_{abc} \epsilon = 0  \eea

We will now look for supergravity solutions which preserve 8 of the
16 background supersymmetries. We will also demand that the complete
solution respect the symmetries of $AdS_2 \times S^2$.

We may then write the complete metric in the form \bea ds^2 = -M(\psi
,\theta) (d\psi ^2 +d\theta^2 ) -N(\psi ,\theta)\cosh^2 \psi
(d\omega^2 -\cosh^2 \omega d\tau ^2)
\nonumber\\
-P(\psi ,\theta)\sin^2 \theta (d\phi^2 +\sin^2 \phi d\chi^2 ) \eea

We have used a diffeomorphism to set the coefficients of $d\psi ^2$
and $d\theta^2$ equal to each other.  This is possible because, by
construction, the unknown functions in the metric only depend on
$\psi,\theta$. We may also perturb all of the NS-NS and RR field
strengths, but the demands of $AdS_2 \times S^2$ symmetry require
that the perturbed field strengths (in tangent space coordinates)
depend only on $\psi$ and $\theta$. Unlike \cite{Kumar:2002wc},
\cite{Kumar:2003xi} we will not assume 6d self-duality of the
three-form field strength.

Since the solution preserves 8 supersymmetries, there must be a
Killing spinor $\epsilon$ which  satisfies $\Gamma \epsilon
=\epsilon$ for some projector $\Gamma$. Given the symmetries of the
problem, the most general projector can be rewritten in the form
$\gamma_{\tau \omega } \epsilon^* =[A(\psi ,\theta) -B(\psi
,\theta)\gamma_{\psi \theta}] \epsilon$. This projector will act on a
Killing spinor of the form \bea \epsilon =[m(\psi ,\theta) +n(\psi
,\theta)\gamma_{\psi \theta}] \tilde \epsilon \eea where $\tilde
\epsilon$ is a Killing spinor of the background and $m$ and $n$ are
complex functions.
We will not specify $A$, $B$, $m$ or $n$, but will leave them as free
functions to be solved for by the ${1\over 2}$-BPS condition.

We may immediately demonstrate, following \cite{Kallosh:1997qw}, that
\bea
Z={e^{\tilde \chi} _{\chi} \over \bar{e}^{\tilde \chi} _{\chi}
}=
P^{-{1\over 2}}
={e^{\tilde \tau} _{\tau} \over \bar{e}^{\tilde \tau} _{\tau} }
=N^{-{1\over 2}} = (mm^* +nn^*)^{-1}.
\eea
This is because translation along the $\tau$
and $\chi$ directions are isometries of the full solution. The
supersymmetry algebra (in the absence of charges) is \bea \{
Q_{\alpha } ^A,\bar Q_{\beta} ^B \} &=& \delta^{AB} P_{\mu}
[(1+\gamma^{11}) \gamma^{\mu} ]_{\alpha \beta} \eea

Thus, the commutator of any two supersymmetries will be a
translation. This is valid in any region of space-time where there
are no charges (so the supersymmetry algebra is not deformed by
central charges).

In particular, since
\bea
\delta_{\epsilon} =\epsilon Q +\bar
\epsilon \bar Q
\eea
we find
\bea
\delta_{\epsilon}
\delta_{\epsilon'} - \delta_{\epsilon'} \delta_{\epsilon} &=&
2\delta^{AB}[\bar \epsilon_A  \gamma^{\mu} \epsilon'_B -
\bar \epsilon'_B \gamma^{\mu}  \epsilon_A ] P_{\mu}
\nonumber\\
&=&
4\imath \delta^{AB}Im [\bar \epsilon_A  \gamma^{\mu} \epsilon'_B ]
P_{\mu}
\eea
where $P_{\mu}$ is the
generator of translations.

A supersymmetry transformation by a Killing spinor will leave the
supergravity fields invariant. Therefore, the commutator of two
supersymmetry transformations by Killing spinors must be a
translation under which the supergravity fields are invariant, ie. an
isometry.  Indeed, one can verify that the isometries of translation
along $\tau$ and $\chi$ of the background are generated by a
commutator of Killing spinor supersymmetry transformations.

The parameter associated with translation along the isometry
direction will thus be related to the Killing spinors by the
supersymmetry algebra.  This translation parameter must be a
constant, since the solution is invariant under constant shifts of
$\tau$ or $\chi$.

In particular, if
$x^{\mu}$ corresponds either of these two isometry directions (one
does not sum over $\mu$), then
\bea
4Im(\bar \epsilon \gamma^{\tilde
\mu} \epsilon') {\partial \over
\partial x^{\tilde \mu} } &=& 4Im(\bar \epsilon \gamma^\mu \epsilon')
e_\mu {}^{\tilde \mu} \; {\partial  \over \partial x^{\tilde \mu} }
\nonumber\\
&=& 4Im(\bar \epsilon_o  \gamma^\mu \epsilon_o ')  (mm^* +nn^*)
e_\mu{}^{\tilde \mu} \; { \partial  \over \partial x^{\tilde \mu} }
\nonumber\\
&=& const \times { \partial  \over \partial  x^{\tilde \mu} } \eea
where in the last step we have set the anti-commutator of Killing
spinors to be a Killing vector.
Note that the background solution must also satisfy this constraint.
Using this and the boundary condition at infinity, we find
\begin{equation}
(mm^* +nn^*) {e_{\mu}{}^{\tilde \mu} \over \bar e_{\mu}{}^{\tilde
\mu} } = 1 \qquad \Longrightarrow \qquad mm^*+nn^* =
({e_{\mu}{}^{\tilde \mu} \over \bar e_{\mu}{}^{\tilde \mu} })^{-1}
\end{equation}

\section{Killing equations}

We can now substitute the ansatz into the Killing equations. The
$\psi$ Killing equation can be written as
\bea
\gamma^\psi{\del_\psi
(m+n\g_{\psi\theta})} \tilde\epsilon -{1\over 2} \omega_{ \psi}
^{\psi \theta} \gamma_{ \theta} \epsilon -{1\over 4} Q_\psi \g_\psi\e
+{\imath \over 8}  f^{\tau\omega \phi \chi \theta} \gamma_{\tau\omega
\phi \chi \theta} \epsilon -{\imath \over 8} f^{\psi\tau \omega \phi
\chi } \gamma_{\psi\tau\omega \phi \chi } \epsilon
\nonumber\\
-{3 \over 8}  f^{\psi\tau \omega} \gamma_{\psi\tau \omega} \epsilon^*
+ {1 \over 8}  f ^{\tau\omega \theta} \gamma_{\tau\omega \theta}
\epsilon^*
-{3 \over 8}  f^{\psi\phi \chi} \gamma_{\psi\phi \chi} \epsilon^* +{1
\over 8}  f^{\phi \chi \theta} \gamma_{\phi \chi \theta}
\epsilon^* ~~~~~~~~~~~~~~~~ \nonumber\\
~~~~~~~~~~=\gamma^\psi{ (m+n\gamma_{\psi\theta})} \left(-{1\over 2}
\bar{\omega}_{ \psi} ^{\psi \theta} \gamma_{\psi \theta}
\tilde\epsilon -{3 \over 8} \bar{ f} _{\psi\tau \omega} \gamma^{\tau
\omega} \tilde\epsilon^* +{1 \over 8} \bar{ f}^{\phi \chi \theta}
\gamma_{\psi \phi \chi \theta}\tilde \epsilon^*\right) {e_{\psi }
^{\tilde \psi }  }
\eea
where we have evaluated derivatives of the background Killing
spinor in terms of the background values of the vielbein and
field strengths.

We let the $a$ index represent one of the directions along the $T^4$.
The Killing equation along this direction is  \bea -{1\over 2}
\omega_{a} ^{a \psi} \gamma_{\psi} \epsilon - {1\over 2} \omega_{a}
^{a \theta} \gamma_{\theta} \epsilon +{\imath \over 8} f^{\psi \tau
\omega \phi \chi} \gamma_{\psi \tau \omega \phi \chi} \epsilon
+{\imath \over 8} f^{\tau \omega \phi \chi \theta} \gamma_{\tau
\omega \phi \chi \theta} \epsilon +{1 \over 8}
 f^{\psi \tau \omega} \gamma_{\psi \tau \omega} \epsilon^*
\nonumber\\
+{1 \over 8}  f^{\psi \phi \chi} \gamma_{\psi \phi \chi} \epsilon^*
+{1 \over 8}  f^{\tau \omega \theta} \gamma_{\tau \omega \theta}
\epsilon^* +{1 \over 8}  f^{\phi \chi \theta} \gamma_{\phi \chi
\theta} \epsilon^* = 0 \eea

The other Killing equations are similar. The spinor now satisfies
$\gamma_{\tau \omega } \epsilon^* =[A(\psi ,\theta) -B(\psi
,\theta)\gamma_{\psi \theta}] \epsilon$. Using this, we can simplify
the Killing equations; for example, we get
\bea \g_{\psi\tau\omega}\e^* &=& -\gamma_{\phi \chi \theta}
\epsilon^* = [A\gamma_{\psi} +B\gamma_{\theta}] \epsilon
\nonumber\\
\gamma_{\tau \omega \theta} \epsilon^* &=& \gamma_{\psi \phi \chi }
\epsilon^* = [A\gamma_{\theta} -B\gamma_{\psi }]\epsilon . \eea

Each Killing equation then produces a coefficient of $\g_\psi\e$ and
$\g_\theta\e$. The coefficients should vanish separately. The Killing
equations therefore split into twelve complex algebraic equations,
which can be used to solve for the various field strengths and for
the vielbein.

The composite connection gauges a local $U(1)$ symmetry under which
the Killing spinor is charged.  We will thus use a gauge
transformation to make $m$ real. The Killing equations in the $a$
direction immediately tell us that $n$ is real and that \bea  A &=& -
{\cosh \psi \over \cosh^2 \psi -\sin^2 \theta} (\sinh \psi +\imath
\cos \theta)
\nonumber\\
 B &=& {\imath \sin \theta \over \cosh^2 \psi -\sin^2 \theta}
(\sinh \psi +\imath \cos \theta) \eea (Note that this form of the
projector was exactly the form found in \cite{Kumar:2002wc} using
$\kappa$- symmetry \cite{Cederwall:1996pv} arguments  for a D3-brane
source wrapping an $AdS_2 \times S^2$ submanifold.)

We will now look for localized solutions in the linearized
approximation. We therefore approximate \bea { e_{\tilde \mu \mu}
\over \bar e_{\tilde \mu \mu}} &=& 1+\d_\mu \eea We can then solve
the Killing equations to find \bea \d_\psi-\d_\omega =
-2(\delta_{\omega} +\delta_a ) =-{c\over \gamma} \eea where $\gamma
=\sin \theta \cosh \psi$ and $c$ is a constant of integration
parameterizing the solution.

We also find \bea  f^{\tau \omega \theta} -f^{\psi \phi \chi} =
{\imath c\over \gamma} ({1\over \gamma} +\imath \tanh \psi \cot
\theta)~~~~~~~~~~~~~~~~~~~~~
\nonumber\\
 f^{\psi \tau \omega } + f^{\phi \chi \theta } = 2Z+{c\over
\gamma}~~~~~~~~~~~~~~~~~~~~~~~~~~~~~~~~~~~~~
\nonumber\\
 f^{\tau \omega \theta} + f^{\psi \phi \chi} = {\imath  A
\over
 A^2 +  B^2} (- f^{\psi \tau \omega \phi \chi} +4\imath
\partial_{\tilde \theta}
\log e^{\tilde a} _a )
-{\imath  B \over  A^2 + B^2} (f^{\tau \omega \phi \chi \theta }
+4\imath \partial_{\tilde \psi} \log e^{\tilde a} _a)
\nonumber\\
 f^{\psi \tau \omega } - f^{\phi \chi \theta } = {\imath  B
\over A^2 +  B^2} (-f^{\psi \tau \omega \phi \chi} +4\imath
\partial_{\tilde \theta} \log e^{\tilde a} _a)
+{\imath  A \over  A^2 + B^2} (f^{\tau \omega \phi \chi \theta }
+4\imath \partial_{\tilde \psi} \log e^{\tilde a} _a)
\nonumber\\
f_{\psi \tau \omega \phi \chi} = \imath Q_{\theta } \qquad f_{\tau
\omega \phi \chi \theta} = -\imath
Q_{\psi}~~~~~~~~~~~~~~~~~~~~~~~~~~~~~~~
\nonumber\\
p_{\tilde \psi} ( A^2 + B^2)  = {\imath \over 2} {1+\beta^2 \over
1-\beta^2 } ( f^{\tau \omega \phi \chi \theta } +4\imath
\partial_{\tilde \psi} \log e^{\tilde a} _a ) +{\beta \over
1-\beta^2} (- f^{\psi \tau \omega \phi \chi} +4\imath
\partial_{\tilde \theta} \log e^{\tilde a} _a)
\nonumber\\
p_{\tilde \theta} ( A^2 +  B^2)  = -{\beta \over 1-\beta^2 } (
f^{\tau \omega \phi \chi \theta } +4\imath
\partial_{\tilde \psi} \log e^{\tilde a} _a )
+{\imath \over 2 }{1+\beta^2 \over 1-\beta^2} (- f^{\psi \tau \omega
\phi \chi} +4\imath \partial_{\tilde \theta} \log e^{\tilde a} _a )
\eea

So far $e^{\tilde a} _a, Q_{\theta }$ and $Q_{\psi}$ are arbitrary
functions. If we demand that all brane sources vanish within any
region, then we immediately find from the equations of motion that
$e^{~\tilde a} _a=1$ and that the 5-form field strength $f_5$
vanishes within this region at linear order. The field strengths then
become self-dual. The analysis is then identical to the analysis of
\cite{Kumar:2002wc}.

Thus, the only solution in which the charge distribution does not
fill space-time (to linear order) is the solution found previously in
\cite{Kumar:2002wc}\footnote{Note that $c$ in this paper is
equivalent to ${1\over c}$ in the notation of \cite{Kumar:2002wc}.}.
The source for this solution is not localized; it is D-string charge
delocalized in the $\psi$ direction (though it is localized in
$\theta$ at $\theta =0$). The charge density was found to be
$\rho_{\tau \omega} = {2\pi c \over \cosh^2 \psi} \delta(\theta )$.
This implies that no solutions with localized charges are possible.

\section{The nonlinear solution}

If we do not restrict ourselves to the linearized approximations, the
solution is more complicated. It is most conveniently written in
terms of the functions:
\bea X= e_{\tilde \psi }^{\psi } e_{\omega
}^{\tilde \omega }  \cosh \psi  =  e_{\tilde \theta }^{\theta
} e_{\omega }^{\tilde \omega } \cosh \psi  \qquad
\qquad\qquad\qquad\qquad
 R = e_{a }^{\tilde a }~~
\nonumber\\
Z = e_{\omega }^{\tilde \omega } \cosh \psi  = e_{\tau
}^{\tilde \tau }  \cosh \psi \cosh \omega = e_{\phi }^{\tilde
\phi }  \sin \theta
= e_{\chi }^{\tilde \chi}  \sin \theta \sin \phi~~~
\eea

We find from the Killing equations that \bea X &=& {\gamma \over
c+\gamma}
\nonumber\\
ZR &=& \sqrt{X}
\nonumber\\
 m^2 &=&  Z^{-1}
\nonumber\\
{\imath \over 2} Q_{\tilde \theta } &=& {1\over 2} {X\over Z} \
f^{\psi \tau \omega \phi \chi}
\nonumber\\
{\imath \over 2} Q_{\tilde \psi } &=& -{1\over 2} {X\over Z} f^{\tau
\omega \phi \chi \theta } \eea

The 3-form field strength can also be solved for: \bea  f^{\psi \tau
\omega } + f^{\phi \chi \theta} &=& Z+{Z\over X}
\nonumber\\
 f^{\tau \omega \theta } - f^{\psi \phi \chi } &=& (Z-{Z\over
X}) (\tanh \psi \cot \theta - {\imath \over \sin \theta \cosh \psi})
\nonumber\\
 f^{\tau \omega \theta} + f^{\psi \phi \chi} &=& {\imath
 A \over  A^2 +  B^2} (-{X\over Z} f^{\psi \tau \omega \phi
\chi} +4\imath \partial_{\tilde \theta} \log R)
\nonumber\\
&-&{\imath  B \over  A^2 +  B^2} ({X\over Z} f^{\tau \omega \phi \chi
\theta } +4\imath \partial_{\tilde \psi} \log R)
\nonumber\\
 f^{\psi \tau \omega } -f^{\phi \chi \theta } &=& {\imath
 B \over  A^2 +  B^2} (-{X\over Z} f^{\psi \tau \omega \phi
\chi} +4\imath \partial_{\tilde \theta} \log R)
\nonumber\\
&+&{\imath  A \over  A^2 +  B^2} ({X\over Z} f^{\tau \omega \phi \chi
\theta } +4\imath \partial_{\tilde \psi} \log R)
\eea

The dilaton-axion field strength can be written as
\bea p_{\tilde \psi} ( A^2
+ B^2)  &=& {\imath \over 2} {1+\beta^2 \over 1-\beta^2 } ({X\over Z}
f^{\tau \omega \phi \chi \theta } +4\imath \partial_{\tilde \psi}
\log R)
\nonumber\\
&+&{\beta \over 1-\beta^2} (-{X\over Z} f^{\psi \tau \omega \phi
\chi} +4\imath \partial_{\tilde \theta} \log R)
\nonumber\\
p_{\tilde \theta} ( A^2 +  B^2)  &=& -{\beta \over 1-\beta^2 }
({X\over Z} f^{\tau \omega \phi \chi \theta } +4\imath
\partial_{\tilde \psi} \log R)
\nonumber\\
&+&{\imath \over 2 }{1+\beta^2 \over 1-\beta^2} (-{X\over Z} f^{\psi
\tau \omega \phi \chi} +4\imath \partial_{\tilde \theta} \log R)
\eea
where $\beta = {\sin \theta \over \cosh \psi}$. Thus, the entire
solution is parameterized by the (arbitrary) functions $R$ and the
5-form field strength $f_{5}$.

Note that we have not yet used the equations of motion to impose
additional constraints; we are allowing the possibility
of space-filling charge-distributions.  The constraints implied
by the equations of motion are more
difficult to impose beyond linear order.  However, if the solutions
without space-filling charges are necessarily identical to those
found in \cite{Kumar:2002wc} at linear
order, it is difficult to see how they could differ at higher
order.

\section{Conclusions and Open Questions}

Our main result is that there are no local states that preserve half
the supersymmetry and have geometry $AdS_2\times S^2$. The analysis
was simplest in the linearized approximation, and this is already
sufficient to show that no such state can exist.

At the nonlinear level, we can still obtain exact results for some
fields (in particular, we can solve for $e_{\tilde \psi }^{~\psi }
e_{\omega}^{~\tilde \omega } $ exactly). Furthermore, if we demand
that there exist a region where sources vanish, then the full
non-linear solution necessarily reduces to the solution we had
previously found in \cite{Kumar:2002wc} which has a D-string source
which is localized in $\theta$, but not $\psi$. It is interesting
that one may find non-localized solutions which preserve 8
supersymmetries, but not solutions for localized branes.

This result may seem quite surprising, since the boundary state
argument of \cite{Bachas:2000fr} implies that such a localized
${1\over 2}$-BPS brane should exist. We believe that the most likely
resolution of this paradox is that the D-brane in fact preserves only
one-quarter of the supersymmetries. (Such solutions have been found
in \cite{Rajaraman:2002vf}). In the classical boundary state
analysis, it appears that half the supersymmetries are preserved, but
interactions must break part of the supersymmetry. This suggests that
the boundary state construction in $AdS_3$ is somewhat subtle. The
true boundary state which satisfies Cardy's condition must presumably
satisfy a deformed version of the boundary condition. This might have
relevance to the problems pointed out in \cite{Ponsot:2002te}.
Understanding these issues would be quite interesting.

Another possibility is that the D-brane source is bent in the full
solution, and that the embedding manifold is thus deformed away from
$AdS_2 \times S^2$. But given the amount of symmetry in the problem,
it is difficult to see how the brane geometry could be deformed away
from $AdS_2 \times S^2$ at all, let alone at leading order.

If solutions for localized ${1\over 2}$-BPS branes had existed, they
would have allowed one to examine the puzzling lack of charge
quantization in spaces such as $AdS_3 \times S^3$ which
asymptotically are curved and have fluxes turned on
\cite{Bachas:2000ik}.
Given the absence of such solutions, the best chance for studying the
issue of Dirac quantization in this space would be to generate
localized ${1\over 4}$-BPS brane solutions.  We hope to return to
this issue in future work.

\vskip .5in

{\bf Acknowledgments}

We are grateful to M. Duff, K. Intriligator and J. Liu for useful
discussions.  The work of J. K. is supported by the Michigan
Center for Theoretical Physics and the
Department of Energy.

\end{document}